# Electrically addressing a single self-assembled quantum dot

D.J.P. Ellis,[a)] A.J. Bennett, and A.J. Shields
*Toshiba Research Europe Limited, Cambridge Research Laboratory, 260 Science Park, Milton Road, Cambridge, CB4 0WE, United Kingdom*

P. Atkinson, and D.A. Ritchie
*Cavendish Laboratory, Cambridge University, Madingley Road, Cambridge,CB3 0HE, United Kingdom*

**Abstract**

We report on the use of an aperture in an aluminum oxide layer to restrict current injection into a single self-assembled InAs quantum dot, from an ensemble of such dots within a large mesa. The insulating aperture is formed through the wet-oxidation of a layer of AlAs. Under photoluminescence we observe that only one quantum dot in the ensemble exhibits a Stark shift, and that the same single dot is visible under electroluminescence. Autocorrelation measurements performed on the electroluminescence confirm that we are observing emission from a single quantum dot.

The quantized energy levels of semiconductor quantum dots (QDs) have facilitated much fundamental physics research and device applications such as single photon sources. For many of these experiments it is necessary to isolate only one dot from an ensemble to probe its quantum nature. Much of the initial work on InAs/GaAs QDs involved the use of micron-sized apertures in an opaque metal film above the ensemble to resolve the area.[1,2] In a particular example, Yuan *et al* used this technique to demonstrate an electrically driven single photon source where the aperture restricted the area of the device from which emission is collected.[3] Also, it is possible to reduce the volume of the active area by etching mesas around the QD. Such structures can also act to confine the light and have been demonstrated in numerous designs including micropillars,[4,5,6] microdisks,[7,8] and photonic crystals.[9,10] However, creating reliable electrical contacts to a submicron-sized mesa presents a considerable technological challenge. A third option is to form an annulus of insulating aluminum oxide (AlOx) within a larger mesa. This is achieved by oxidizing an aluminum-rich $Al_xGa_{1-x}As$ layer at elevated temperatures and in a humid environment. It has been demonstrated that, through careful choice of $Al_xGa_{1-x}As$ composition and wet-oxidation conditions, apertures of diameter down to around 100nm can be produced.[11]

In this letter, we describe a device in which an aperture in an AlOx layer, formed by wet oxidation of an AlAs layer, has been used to isolate a single quantum dot within a p-i-n LED structure. We show that despite the presence of a number of QDs within the device, we are able to electrically address a single dot by this method.

The structures have been processed from wafers grown by molecular beam epitaxy. Self assembled InAs QDs (2.1ML grown at 485ºC) were placed in a 700nm layer of undoped GaAs, situated between n- and p-doped layers of GaAs. A 40nm thick layer of AlAs was also located within the undoped region, 20nm from the QD layer. Mesas ranging in size up to 50µm were defined by standard photolithography and wet etching. The current apertures were then formed by placing the samples in a tube furnace, heated to 400ºC. The moist environment was created by passing $N_2$ gas through de-ionised water, held at 98ºC, into the furnace tube. Oxidation proceeds laterally into the structures through the edges of the AlAs layers which were exposed as the mesas were etched. By varying the oxidation time, it was possible to control the diameter of the unoxidised region in the centre of the device. An optical microscope was used to directly measure the oxidation length of a control sample, also containing a 40nm AlAs layer, which was oxidized simultaneously. These observations allowed us to estimate the oxidation rate to be (1.1±0.1)µm/minute, and the diameter of the aperture in our sample to be less than 1.5µm.

An insulating layer of silicon nitride was deposited, by plasma-enhanced chemical-vapor deposition, and subsequently patterned prior to the deposition of ohmic contact material and a final TiAu metallization. The materials used for the ohmic contacts were carefully chosen so that the associated rapid thermal anneal (RTA) steps would not result in device delamination at the oxide interface – a problem which plagued early attempts to integrate wet-oxidation technology into device structures.[12] PdGe was employed to form a shallow contact to the n-GaAs side of our device and indium tin oxide (ITO) was used for the p-contact. A schematic of our device is shown in Fig. 1.

Characterization of the device was carried out in a liquid-He cooled continuous-flow optical cryostat. Direct current (dc) electrical characterization of our devices show they have a diode-type current-voltage characteristic, with a threshold voltage ($V_T$) of 2.03V, above which QD electroluminescence (EL) was observed. The threshold voltages measured for all such devices are consistently around 0.55V higher than expected for a p-i-n device. We believe this is due to high contact resistance.

Photoluminescence (PL) characterization was carried out with the sample excited at normal incidence using a 770nm picosecond-pulsed laser. The emission was collected with the same microscope objective and analyzed with a spectrometer and a liquid-nitrogen-cooled CCD. Spatial mapping of the PL emission from the device, generated by stepping the microscope objective in a line across the middle of the device, revealed the presence of many individual QDs with emission lines around 900nm. Fig. 2(a) shows spatially resolved PL from near the centre of the mesa. Here we observe three distinct sets of emission lines within about 4μm, attributed to three different dots (labeled A, B, and C).

The PL spectra obtained from a large area of the device under different bias conditions below threshold is shown in Fig. 2(b). It reveals two distinct groups of lines – those where the emission wavelength is invariant with applied voltage and those where a change is observed. By comparing the spectra with the spatially mapped PL, we were able to identify that the lines with constant emission wavelength originate from dots A and C, as indicated in the figure. At ~1.8 – 2.0V we also observe a pair of lines that were previously identified as emission from dot B. As the bias is reduced, the wavelength of these lines decreases to a minimum near 1.5V. Around this point there is a marked changed in the spectra: these lines extinguish and three other features appear at longer wavelengths. As the bias is further reduced, these lines also exhibit a Stark shift. This suggests that the five lines may correspond to emission from different neutral and charged states of a single QD. Power dependence studies reveal that three of the lines have a linear dependence whilst the remaining two exhibit quadratic behaviour. The lines therefore correspond to excitonic and biexcitonic transitions respectively. The linear polarization of the lines was then studied. Only one pair of lines exhibit a wavelength splitting between horizontally and vertically polarized states. Such a splitting would be expected between X and $X_2$ states, but not between charged states at zero magnetic field. By combining these observations, we are able to determine that the five emission lines correspond to the X, $X_2$, $X^+$, $X_2^+$, and $X^-$ transitions of a single QD, as indicated in Fig. 2(b). Such a characteristic line structure is often observed from QDs at these wavelengths.[13,14] Furthermore, since the emission from dots A and C shows no wavelength shift with applied voltage, we must conclude that the electric field they experience does not change. This would suggest that only QD B is located below our aperture region. From our spatial mapping we are able to assess the spacing between dots A,B and C and hence estimate our aperture to have a diameter of around 1µm.

In contrast to the PL, the composition of the EL spectra [Fig. 2(c)] obtained from the device did not change as the microscope objective was scanned across the sample. We observe three clear features: the $X^+$ and $X_2^+$ lines of the QD under the aperture (dot

B), together with a line from a nearby QD (dot A), most likely on the periphery of the current path and hence weak in intensity. Fig. 2(d) plots the integrated intensity of the two bright lines as a function of current. This data indicates that the 896.0nm and 898.1nm line have linear and quadratic current dependencies and confirms our assignment of $X^+$ and $X_2^+$ states to these lines.

To prove that we are able to isolate the EL of a single dot using the emission aperture, we studied the statistics of photon emission by performing autocorrelation measurements using a free-space Hanbury-Brown and Twiss interferometer consisting of a beam-splitter, two avalanche photo-diodes and time-correlated counting electronics. Fig. 3(a) shows an autocorrelation histogram obtained under dc electrical injection. We see a clear dip to around 0.4 at zero time delay, indicating the anti-bunching of emitted photons. The data has been modeled using a three-level rate-equation model, similar to that previously discussed by Shields *et al.*[15] Transitions are taken to be purely radiative and the finite time response of our system, together with the known background emission levels are also taken into account. The agreement with the experimental data is excellent for the $X^+$ state, with a radiative lifetime of 450ps.

We repeated this experiment using electrical pulses, of nominal length 300ps and at a repetition rate of 80MHz, to drive the device. Data collected from the $X^+$ state is presented in Fig. 3(b) for a small time-averaged current of 10nA. The reduction in height of the zero-delay peak, to below 0.5, clearly indicates non-classical emission. Spectral measurements suggest the area of central peak should be 24% if the signal is due entirely to a perfect single-photon-emitting emission line superimposed on a spectrally flat background emission with classical photon statistics. In order to operate efficiently under pulsed electrical operation, we require electrical pulses of width less than the lifetime of our dot (450ps). However, the impedance mismatch between the 50Ω pulse generator and our device results in temporal broadening of the electrical injection pulse to ~ 400ps. This in turn increases the probability that the dot may "refill" after the emission of an initial photon. This is likely to be the source of the remainder of the area of the zero-delay peak. Optimization of both the electrical structure and any impedance matching networks required would, when combined with shorter pulses, reduce this effect.

We have demonstrated that an aperture in an aluminum oxide layer, formed through wet oxidation can be used to electrically isolate and address a single self-assembled InAs quantum dot within a large device. This technique may be useful for a wide range of devices, including single dot photodiodes, electrically driven single photon sources and gated schemes for QD based quantum computing. We note that the format of the oxide aperture is compatible with the incorporation of cavity structures to enhance the photon-dot coupling efficiency.[16,17,18,19]

**Acknowledgements**

This work was partially supported by the European Commission under

Framework Package 6 Network of Excellence SANDiE and the Integrated

Project Qubit Applications (QAP) funded by the IST directorate as Contract Number 015848. One of the authors (D.J.P.E.) would like to thank EPSRC and Toshiba for funding

**Footnotes**

a) Also at: Cavendish Laboratory, Cambridge University, Madingley Road, Cambridge,CB3 0HE, United Kingdom

**Figure Captions**

FIG.1. Schematic of the device structure. A single self-assembled quantum dot is isolated within an aperture of aluminum oxide. The simple RC network is placed in parallel with the device and the power supply in order to reduce the impedance mismatch of the system

FIG. 2. (a) Photoluminescence obtained as the microscope objective was scanned across the centre of the device at a constant bias of 2.0V. Emission was observed from three different quantum dots, labeled A, B, and C. (b) Photoluminescence collected over a large area plotted as a function of voltage for biases below the threshold voltage, $V_T$. Emission lines from quantum dot 'B' are labeled with their electronic configuration. (c) Electroluminescence collected for voltages greater than $V_T$. (d) Integrated electroluminescence intensity of the $X^+$ and $X_2^+$ emission lines as a function of drive current. All figures display the measured intensity on a logarithmic scale.

FIG. 3. Autocorrelations recorded from $X^+$ emission line (a) Experiment performed under dc electrical injection at 2.2V. The solid line is fitted. See text for details. (b) Data recorded under pulsed electrical injection at a repetition rate of 80MHz.

**Fig 1.**

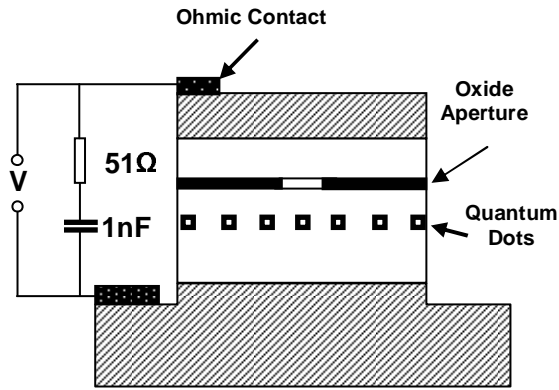

**Fig 2.**

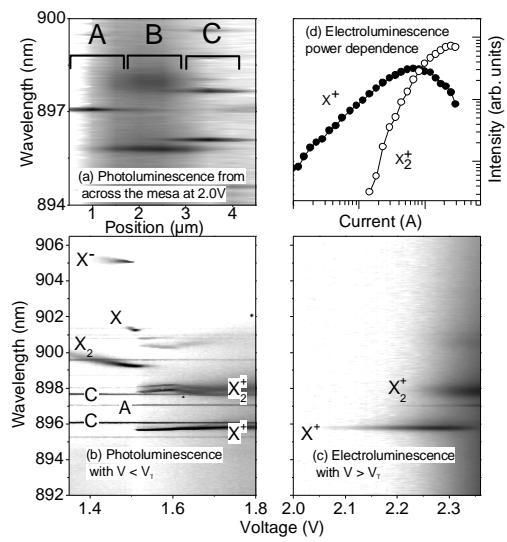

**Fig 3.**

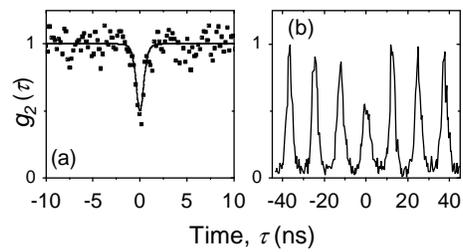